\documentclass[12pt, preprint]{aastex}
\usepackage{bm}

\shorttitle{Mid-IR Exoplanets}
\shortauthors{Skemer et al.}

\begin{document}

\title{Directly Imaged L-T Transition Exoplanets in the Mid-Infrared
\footnote{The LBT is an international collaboration among institutions in the United States, Italy and Germany. LBT Corporation partners are: The University of Arizona on behalf of the Arizona university system; Istituto Nazionale di AstroÞsica, Italy; LBT Beteiligungsgesellschaft, Germany, representing the Max-Planck Society, the Astrophysical Institute Potsdam, and Heidelberg University; The Ohio State University, and The Research Corporation, on behalf of The University of Notre Dame, University of Minnesota and University of Virginia.}
\footnote{This paper includes data gathered with the 6.5 meter Magellan Telescopes located at Las Campanas Observatory, Chile.}
}
\author{Andrew J. Skemer$^{1}$,
Mark S. Marley$^{2}$,
Philip M. Hinz$^{1}$,
Katie M. Morzinski$^{1}$,
Michael F. Skrutskie$^{3}$
Jarron M. Leisenring$^{1,4}$
Laird M. Close$^{1}$,
Didier Saumon$^{5}$,
Vanessa P. Bailey$^{1}$,
Runa Briguglio$^{6}$,
Denis Defrere$^{1}$,
Simone Esposito$^{6}$,
Katherine B. Follette$^{1}$,
John M. Hill$^{7}$,
Jared R. Males$^{1}$,
Alfio Puglisi$^{6}$,
Timothy J. Rodigas$^{1,8}$,
Marco Xompero$^{6}$
}

\affil{$^{1}$Steward Observatory, Department of Astronomy, University of Arizona, 933 N. Cherry Ave, Tucson, AZ 85721}
\affil{$^{2}$NASA Ames Research Center, MS-245-3, Moffett Field, CA 94035}
\affil{$^{3}$Department of Astronomy, University of Virginia, 530 McCormick Road, Charlottesville, VA 22904}
\affil{$^{4}$Institute for Astronomy, ETH Zurich, Wolfgang-Pauli-Strasse 27, CH-8093 Zurich, Switzerland}
\affil{$^{5}$Los Alamos National Laboratory, Mail Stop F663, Los Alamos, NM 87545}
\affil{$^{6}$Istituto Nazionale di Astrofisica, Osservatorio Astrofisico di Arcetri Largo E. Fermi 5 50125 Firenze, Italy}
\affil{$^{7}$Large Binocular Telescope Observatory, University of Arizona, 933 N. Cherry Ave, Tucson, AZ 85721}
\affil{$^{8}$Department of Terrestrial Magnetism, Carnegie Institute of Washington, 5241 Broad Branch Road NW, Washington, DC 20015}

\begin{abstract}
Gas-giant planets emit a large fraction of their light in the mid-infrared ($\gtrsim$3$\micron$), where photometry and spectroscopy are critical to our understanding of the bulk properties of extrasolar planets.  Of particular importance are the L and M-band atmospheric windows (3-5$\micron$), which are the longest wavelengths currently accessible to ground-based, high-contrast imagers.  We present binocular LBT AO images of the HR 8799 planetary system in six narrow-band filters from 3-4$\micron$, and a Magellan AO image of the 2M1207 planetary system in a broader 3.3$\micron$ band.  These systems encompass the five known exoplanets with luminosities consistent with L$\rightarrow$T transition brown dwarfs.  Our results show that the exoplanets are brighter and have shallower spectral slopes than equivalent temperature brown dwarfs in a wavelength range that contains the methane fundamental absorption feature (spanned by the narrowband filters and encompassed by the broader 3.3$\micron$ filter).  For 2M1207 b, we find that thick clouds and non-equilibrium chemistry caused by vertical mixing can explain the object's appearance.  For the HR 8799 planets, we present new models that suggest the atmospheres must have patchy clouds, along with non-equilibrium chemistry.  Together, the presence of a heterogeneous surface and vertical mixing presents a picture of dynamic planetary atmospheres in which both horizontal and vertical motions influence the chemical and condensate profiles.
\end{abstract}

\section{Introduction\label{Introduction}}
It is generally assumed that the atmospheres of gas giant planets are analogous to the atmospheres of old brown dwarfs because they are (1) approximately the same composition, (2) approximately the same radius ($\sim$1 R$_{jup}$ supported by electron degeneracy pressure) and (3) slowly cooling due to a lack of internal fusion.  However, early studies of the handful of directly imaged exoplanets suggest that planets may be cloudier and more turbulent than their older and more massive analogs \citep{2008Sci...322.1348M,2010ApJ...723..850B,2011ApJ...729..128C,2011ApJ...733...65B,2011ApJ...732..107S,2011ApJ...735L..39B,2013Sci...339.1398K}.  A particularly large discrepancy exists in the L-band (3-4$\micron$), where atmospheric models that have been successfully used to fit brown dwarfs underpredict the [3.3$\micron$] fluxes of the HR 8799 planets by 1-2 magnitudes \citep{2012ApJ...753...14S}.

\begin{figure}
\begin{center}
\includegraphics[angle=90,width=\columnwidth]{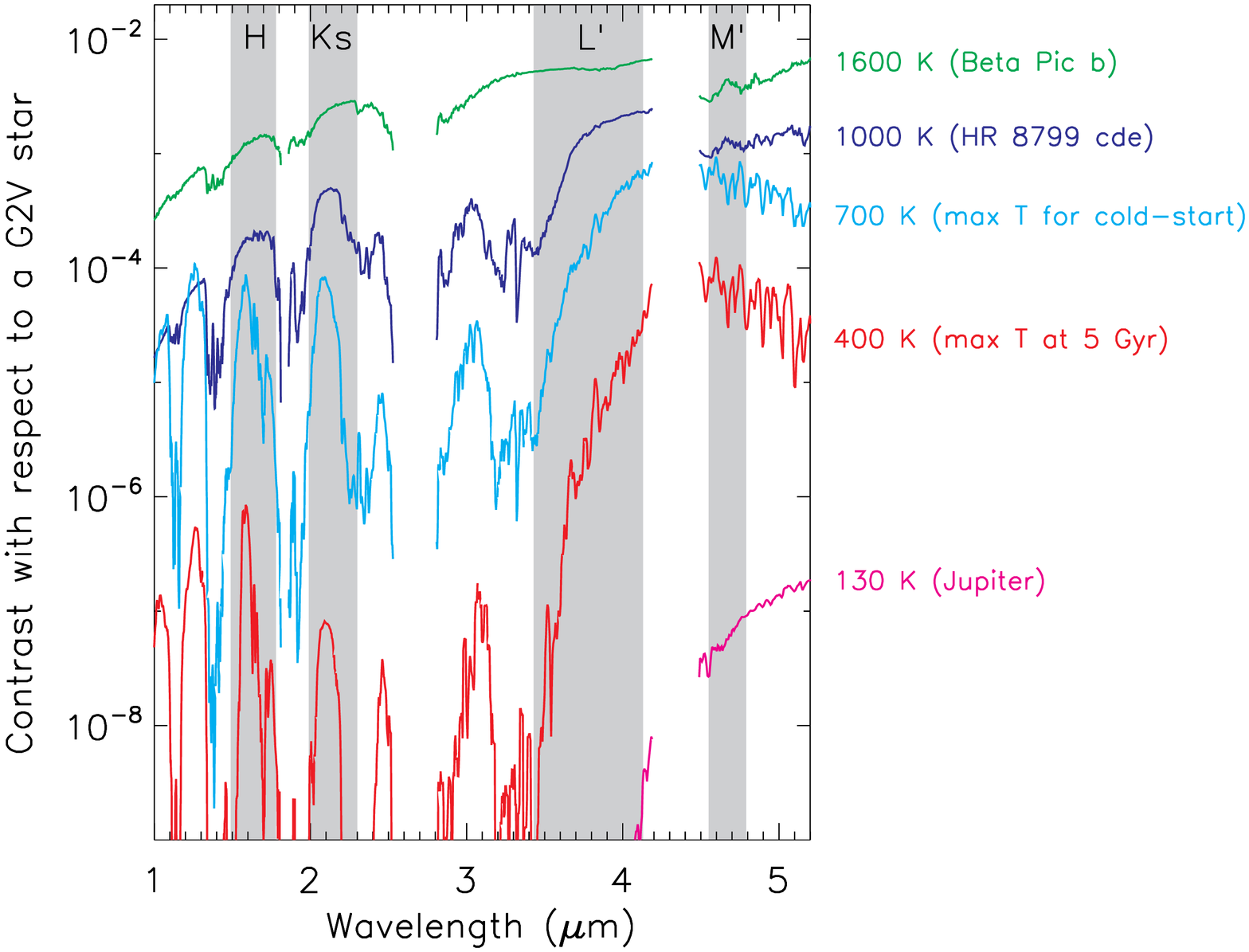}
\vspace{0.1in}
\caption{Characteristic examples of exoplanet-to-star contrasts (i.e. flux ratios) as a function of wavelength, showing (1) that gas-giant exoplanets can be detected with lower contrasts in the mid-infrared (3-5$\micron$) than in the near-infrared (1-2$\micron$), and (2) that this difference increases at lower temperatures.  While the planets that have been directly imaged to date ($\beta$ Pic b and HR 8799 c, d and e on this plot) are relatively warm (1600 K and 1000 K, respectively), it is likely that the majority of self-luminous exoplanets are much cooler.  Planets that formed by core-accretion \citep[approximated by the ``cold-start'' models;][]{2008ApJ...683.1104F} are never hotter than $\sim$700 K.  Planets around average-aged stars (5 Gyr) are never hotter than $\sim$400 K, regardless of formation history.  Jupiter, which may be a ubiquitous outcome of planet formation, is only $\sim$130 K. \citep[Models from ][]{2011ApJ...737...34M,2003ApJ...596..587B}.
\label{mid-IR examples}}
\end{center}
\end{figure}

Much of what is currently known about the physical properties of exoplanets comes from near-infrared (1-2.5$\micron$) photometry and spectroscopy \citep{2004A&A...425L..29C,2007ApJ...657.1064M,2008Sci...322.1348M,2010Natur.468.1080M,2011ApJ...729..128C,2012ApJ...753...14S,2010A&A...517A..76P,2010ApJ...723..850B,2011ApJ...733...65B,2013ApJ...768...24O,2013Sci...339.1398K,2011A&A...528L..15B,2013A&A...555A.107B,2013arXiv1306.0610C,2013ApJ...774...11K,2013arXiv1310.4183J}, due in part to the difficulty of working at longer wavelengths from the ground.  However, self-luminous exoplanets emit the majority of their photons in the mid-infrared $>$3$\micron$, which besides making the region interesting in its own right, implies that less contrast is necessary to detect them against the bright glare of their host stars \citep[see Figure \ref{mid-IR examples} and][]{2006ApJ...653.1486H}.

While the first generation of directly-imaged exoplanets were relatively warm and massive \citep[2M1207 b, HR 8799 bcde and $\beta$ Pic b are all $>$800 K and $>$5 M$_{jup}$;][]{2011ApJ...732..107S,2008Sci...322.1348M,2010Natur.468.1080M,2011A&A...528L..15B}, systems like these are known to be rare, while low-mass, close-in planets are ubiquitous \citep{2013ApJ...773..179W,2013arXiv1306.1233N,2013arXiv1309.1462B,2011ApJ...736...19B,2013ApJS..204...24B}.  Low-mass planets, core-accretion planets, and elderly planets will all be cooler than the early exoplanet discoveries \citep{2008ApJ...683.1104F}.  And ``cool'' planets emit an even larger fraction of their light in the mid-infrared (see Figure \ref{mid-IR examples}).

To prepare for imaging ``cool'' planets, and to understand the ``warm'' planets that have already been found, it is critical that we expand our knowledge of exoplanet SEDs into the mid-infrared.  In this paper, we present deformable secondary AO\footnote{These adaptive optics systems employ the minimum number of warm optics, minimizing the thermal infrared background from the telescope \citep{2000PASP..112..264L}.} imaging of the HR 8799 system, using LBTI/LMIRCam, and the 2M1207 system, using MagAO/Clio2.  HR 8799 is a 20-160 Myr A5V star \citep{1969AJ.....74..375C,2006ApJ...644..525M,2008Sci...322.1348M,2010ApJ...716..417H,2011ApJ...732...61Z} with four directly imaged planets \citep{2008Sci...322.1348M,2010Natur.468.1080M}, whose masses and separations are difficult to explain with standard planet formation models \citep{2009ApJ...707...79D,2010ApJ...710.1375K}.  2M1207 is a 5-13 Myr M8 brown dwarf \citep{2002ApJ...575..484G} with a planetary-mass companion \citep{2004A&A...425L..29C} that is also difficult to explain with standard planet-formation models \citep{2005A&A...438L..25C}.  Ignoring system architectures, HR 8799 bcde and 2M1207 b are unambiguously low-gravity, planetary-mass objects, whose atmospheres offer the first opportunities to characterize directly-imaged planets.  And since all five planets have luminosities consistent with L$\rightarrow$T transition brown dwarfs, their atmospheres are ideal laboratories for studying how cloud properties are affected by a low-gravity, planetary environment.

For HR 8799 we build upon previous [3.3$\micron$] imaging \citep{2012ApJ...753...14S} by using 6 narrow-band filters in the 3-4$\micron$ window to probe the spectral shape of the 3.3$\micron$ methane fundamental absorption feature.  For 2M1207 b, we present photometry in the broader [3.3$\micron$] filter to determine if the object has the same extreme 3.3-3.8$\micron$ colors first seen in the HR 8799 planets.  We present our observations, reductions, and photometry in Section 2, a comparison with field brown dwarfs in Section 3, SED modeling in Section 4, and our conclusions in Section 5.  We also include filter curves and their tabulated properties in Appendix A.

\section{Observations and Reductions\label{Observations and Reductions}}
\subsection{HR 8799\label{HR 8799}}
We observed the HR 8799 planetary system in six 5\% bandwidth filters from 3.0-3.8$\micron$ (see filter properties in Appendix A) on UT Nov. 2, 2012 using LBTI \citep{2012SPIE.8445E..0UH} and its 1-5$\micron$ camera LMIRCam \citep{2010SPIE.7735E.118S,2012SPIE.8446E..4FL}.  The LMIRCam detector is a Hawaii-2RG, 5$\micron$-doped HgCdTe device, with 32 readouts.  Currently we use 16 readouts, each of which reads 1024 rows by 64 columns, giving us a usable region of 1024x1024 (11x11'').  Images from both LBT telescopes were corrected by the LBT's two deformable secondary adaptive optics systems\footnote{The LBTI wavefront sensors are functionally equivalent to the First Light Adaptive Optics Sytem \citep[FLAO][]{2010SPIE.7736E...7E}} and incoherently overlapped in LBTI's beam-combiner. The resulting images at the focal plane of LMIRCam achieved the diffraction-limited performance of a single LBT 8.4 meter aperture and the collecting area of the full 2$\times$8.4 meter LBT. 

Ground-based high-contrast observations are usually limited by instrumental quasi-static speckles, which can be removed by allowing the astronomical field to rotate with parallactic angle, while keeping the instrument rotation fixed \citep[][angular differential imaging]{2006ApJ...641..556M}.  In this approach, sky-rotation and clock-time are the observational requirements, rather than integration time.  For our observations of HR 8799, we switched filters every $\sim$60 seconds, rotating through the set and nodding every $\sim$10 minutes.  By doing this, we were able to achieve adequate sky-rotation and clock-time in multiple filters simultaneously.  We acquired seven minutes of data in each filter (42 minutes total) over a period of 2 hours (with a 1 hour gap due to a telescope malfunction), during which time the parallactic angle changed by 70 degrees.  There were occasional scattered clouds during the night, but the adaptive optics wavefront sensor counts and thermal sky-background stayed consistent throughout the observations.  The LBT's differential image motion monitor (DIMM) measured a natural seeing of 0.8" and the nearby Submillimeter Telescope Tau-meter measured a precipitable water vapor of $\sim$5-6mm.  Our images were taken in correlated double sampling mode (reset-read-integrate-read) so that the first read (0.029 seconds) could be used as an unsaturated image of the star, while the second read (12.1, 12.1, 6.1, 4.0, 6.1, 3.0 seconds for the six filters respectively) saturated the star and filled the wells into the photon-noise regime away from the star.

\subsubsection{Detector Non-Linearity\label{Non-Linearity}}
Throughout Fall 2012, LMIRCam suffered from detector non-linearity, caused by an incorrectly set bias voltage, which has since been re-tuned.  We constructed fluence-to-count calibration curves (linearity curves) by taking sky flats of varying integration times.  Typical detector non-linearity is characterized by decreasing gain with well filling.  LMIRCam's non-linearity had an S-shape where gain increased with well-filling before entering a linear regime and then turning over as the array reached saturation.  Over the course of the semester, the detector linearity did change slightly, but from night-to-night, it was quite consistent.  For the data reduction in this paper, we combined linearity from consecutive nights (UT Nov. 2, 2012 and UT Nov. 3 2012) to improve sampling, allowing an overall change in sky flux as a free parameter.  The pixel-to-pixel linearity curves were mostly consistent (to within $\sim$5\%) after subtracting biases and correcting for flat-fielding effects.  Pixels with linearity curves that were inconsistent with the rest of the array were classified as bad pixels (10\% of the array but mostly concentrated in a region that we avoided during our observations).  Using the good pixels, we constructed a median linearity curve and scaled it, for each pixel, based on a flat-field.  We used the set of curves to pre-process both correlated double sampling reads of every frame, individually.

After reducing all of our data (see Section \ref{LBTI Reductions and Photometry}), the difference in the inferred planet photometry with and without the linearity correction is $\sim$40\%.  To assess the quality of this correction, we took images of HR 8799 with various exposure times, ranging from our minimum integration time to the point where the stellar core began to saturate.  After linearizing these data, the measured stellar flux remained constant to within 3\%, which validates our linearity correction.

\subsubsection{Reductions and Photometry\label{LBTI Reductions and Photometry}}
Our science data is comprised of alternating short and long exposures (correlated double sampling) in a series of six filters.  For the short-integration images, we subtracted similar integration dark frames, taken at the end of the night, to remove pixel-to-pixel biases (which persist for long time-scales).  We then removed the more rapidly changing residual channel biases (which are constant over each readout) by subtracting the median of the 64x8 pixel overscan region from each pixel in the corresponding 64x1016 light-sensitive region.  We then repeated the bias subtraction steps on the long exposure images.  Both sets of images were linearized and flattened.  We then nod-subtracted the images, and subtracted the sigma-clipped median of each column to remove any residual electronic effects.  The images were binned, 2x2, to suppress bad pixels.  Known bad pixels (from flat-field maps) were masked in the process and 2x2 blocks of bad pixels were replaced with the average of surrounding pixels.  The binned pixels have a plate scale of 0.0214''/pixel, which still over-samples the LBT's single-aperture diffraction-limited PSF (0.07'' FWHM) at our shortest wavelength (3$\micron$).  The short and long integration images were registered independently by cross-correlation in Fourier space.  The unsaturated (star) images were median combined with $3\sigma$ clipping. The saturated images were processed for angular differential imaging (ADI).  For each image, the stellar centroid was determined from the Airy rings, and confirmed to be accurate by reducing the first half and second half of our data independently and verifying that the planets did not move.  We then subtracted the stellar profile by fitting an azimuthal profile at every radius bin with the first four low-order modes of a fast-Fourier transform.  Next, we subtracted the median of the stack of images with adequate sky rotation (a parallactic angle that has changed $>\lambda/D)$ at a given radius).  Finally, we rotated and combined the images with a $3\sigma$ clipped median.  

\begin{figure}
\begin{center}
\includegraphics[angle=0,width=\columnwidth]{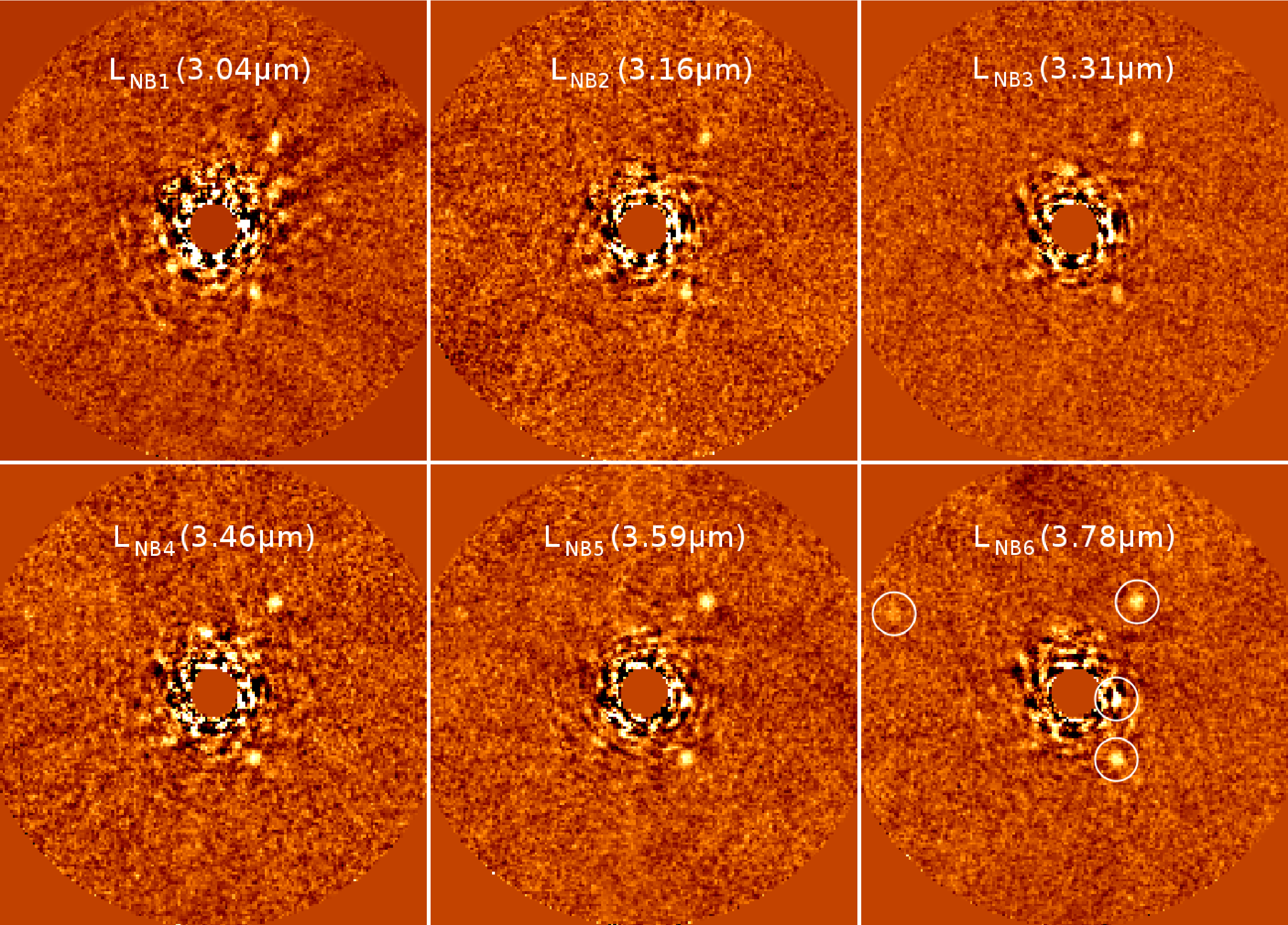}
\caption{LBTI/LMIRCam images of the HR 8799 planetary system in 6 narrowband filters.  Each wavelength was observed for 7 minutes with the incoherently combined 2$\times$8.4 meter LBT aperture.  HR 8799 c and d are visible in all 6 filters.  HR 8799 b is visible in 4 of the 6 filters, although it is low S/N, given the short integration times.  HR 8799 e is also detectable at low S/N in the longest wavelength filters, given its known position.  The planet positions are circled in the lower-right panel. North is up and East is left.  The images are each 3.5'' across. 
\label{HR8799_image}}
\end{center}
\end{figure}

The final reduced images are shown in Figure \ref{HR8799_image}.  Planets c and d are easily detected in all six images.  Planet b is detectable in four of the six images, given its known position, although its S/N is insufficient for detailed photometric modeling.  The noise in the vicinity of planet e is too high to detect the planet in individual narrow-band images (it is detected when the images are coadded).  It is possible planet e could be detected in individual narrow-band images using more sophisticated high-contrast algorithms, such as LOCI or PCA.  However, the detections would not produce photometry that is adequate for this paper's analysis.

We calibrated the photometry of planets c and d by adding artificial negative planets to the positions of the planets before the ADI step \citep[as described in ][]{2012ApJ...753...14S}.  This is necessary to calibrate out the self-subtraction that is intrinsic to high-contrast imaging pipelines, such as ADI.  Our absolute photometry, assuming a stellar apparent magnitude of 5.22 \citep[all bands;][]{2010ApJ...716..417H} and a distance modulus of 2.98$\pm$0.06\citep{2007A&A...474..653V}, is presented in Table \ref{LBT/Magellan}.  Errors were determined empirically from the areas surrounding the planets by inserting artificial negative planets with a range of brightnesses at a precision of 0.05 mags.  The values obtained are consistent with the broad-band measurements of \citet{2008Sci...322.1348M} and \citet{2012ApJ...753...14S} (see Tables \ref{LBT/Magellan} and  \ref{lit phot}).

\subsection{2M1207\label{2M1207}}
We observed 2MASS J12073346-3932539 (2M1207) in a 12\% bandwidth 3.3$\micron$ (see filter information in Appendix A) with the Magellan adaptive optics system \citep[MagAO;][]{2012SPIE.8447E..0XC} and its 14x28'' FOV near-infrared imager, Clio2 \citep{2006SPIE.6269E..27S,2004SPIE.5492.1561F} on UT April 14, 2013.  Conditions were photometric, with 1.0'' seeing (from DIMM measurements) and 4.3 mm precipitable water vapor (from the nearby and similar altitude La Silla weather station).

Previous AO imaging of 2M1207 \citep{2004A&A...425L..29C,2005A&A...438L..25C,2007ApJ...657.1064M} has used VLT/NACO's infrared wavefront sensor \citep{2003SPIE.4839..140R}, which takes advantage of 2M1207 A's very red (V-K=7) color.  The Magellan AO system has a visible light pyramid wavefront sensor \citep{2011SPIE.8149E...1E}, which is able to lock on fainter targets than most typical Shack-Hartmann systems; however, it cannot lock on 2M1207 itself (R$\sim$19.3).  As a result, we locked on a nearby guide-star, 2MASS J12073100-3932281, which is 38.4'' offaxis and has an R magnitude of $\sim$14.2.  The resulting off-axis correction was sufficient to resolve the $\sim$0.8'' binary, 2M1207 A-b.

Clio2 was mounted on the Nasmyth port of Magellan-II (Clay), with the rotator turned on to keep 2M1207 from moving with respect to the off-axis guide star.  We obtained 529 [3.3$\micron$] images of 2M1207 with 1.5 second detector integration times and 4 coadds, totaling 53 minutes of integration time over a period of 2 hours.  The inefficiencies were caused by concurrent engineering work, since this was the first time the system had been used to track an off-axis target through transit at such a high elevation.  Clio2 was configured in its coarse plate-scale mode of 27 mas/pixel.  We nodded back-and-forth by 3'' in an ABBA pattern to subtract the background, and dithered around the chip to aid the removal of bad pixels.

\subsubsection{Reductions and Photometry\label{MagAO Reductions and Photometry}\label{magaoreductions}}
\begin{figure}
\begin{center}
\includegraphics[angle=0,width=0.5\columnwidth]{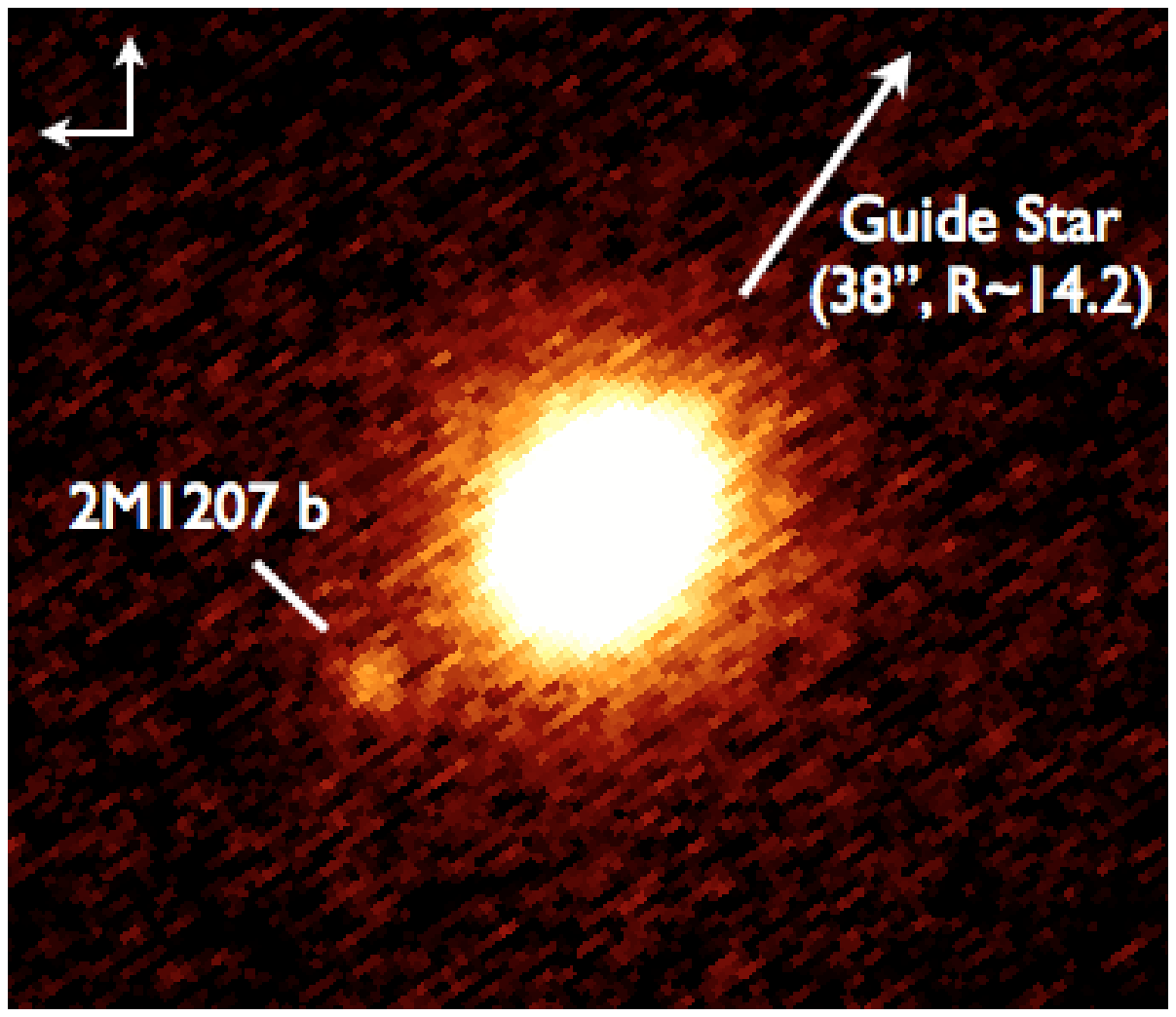}
\caption{MagAO/Clio2 images of the 2M1207 system. The binary is clearly resolved despite being 38'' off-axis from its faint (R$\sim$14.2) AO guide-star.  The slight elongation of 2M1207 A  is from the anisoplanatism of the off-axis adaptive optics reference star.  2M1207 b appears to be circular because of the image stretch, which just shows its diffraction-limited core (FWHM=0.1'').  North is up and East is left.  The field-of-view shown here is 2 by 2.5 arcseconds.
\label{2M1207_image}}
\end{center}
\end{figure}

We processed the data by flat-fielding, interpolating over bad pixels, nod-subtracting the frames, and registering the images.  We frame-selected the best 80\% of the images based on cross-correlation to remove frames with poor image quality.  We then averaged the frames together, which produced a higher S/N detection of 2M1207 b than a median combination. The final image is shown in Figure \ref{2M1207_image}.  2M1207 A and b are clearly resolved.

To cleanly separate 2M1207 A and b for photometry, we subtracted a 5-pixel gaussian smoothed image of the binary from itself (unsharp masking).  This adequately removed 2M1207 A's halo from 2M1207 b's position leaving `b' on a flat background.  We used the IDL Astronomy User's Library's \textit{DAOPHOT-Type} procedures\footnote{http://idlastro.gsfc.nasa.gov/} to perform PSF-fitting photometry, using 2M1207 A as a PSF for 2M1207 b (both unsharp masked).  The resulting PSF-subtracted image had a small negative residual ($\sim$0.2 mags), which we calibrated, along with our statistical uncertainty, using aperture photometry.  For the aperture photometry, we used a 1.5$\lambda$/D aperture on-source (which was background-noise limited after unsharp masking) and around the array to determine statistical errors.  We find that 2M1207 b is 3.86$\pm$0.10 mags fainter than 2M1207 A in the [3.3$\micron$] filter. Our hybrid PSF-fitting/aperture photometry approach was meant to ensure that an elongated PSF (due to anisoplanicity) did not bias our aperture photometry.  However, we note that subtracting the image from itself after a 180 degree rotation (which removes 2M1207 A but not b), and then doing aperture photometry produces a result that is consistent at 1$\sigma$.

Because Clio2's 3.3$\micron$ filter is very sensitive to variable telluric water vapor, we did our absolute calibration using data from the space-based Wide-field Infrared Survey Explorer \citep[WISE;][]{2010AJ....140.1868W}.  In the WISE1 filter (3.4$\micron$), 2M1207 is 11.556$\pm$0.023 mags.  Subtracting the contribution of 2M1207 b ($\Delta$mag$\sim$3.86 at 3.3$\micron$, 2M1207 A is 11.59$\pm$0.02 mags in the WISE1 filter.

To convert the WISE1 photometry to our [3.3$\micron$] bandpass, we calculated the [3.3$\micron$]-WISE1 colors\footnote{We used the 3.3$\micron$ filter curve from the manufacturer (JDSU; see Appendix A), the WISE1 filter curve from http://www.astro.ucla.edu/\textasciitilde wright/WISE/passbands.html, a model Alpha Lyr spectrum from \citet{1992AJ....104.1650C}, and a telluric transmission function (1.0 airmasses, 4.3 mm precipitable water vapor at Cerro Pachon) from http://www.gemini.edu/sciops/telescopes-and-sites/observing-condition-constraints/ir-transmission-spectra.} of a range of DUSTY model atmospheres \citep{2001ApJ...556..357A} that have been used to fit 2M1207 A \citep{2007ApJ...657.1064M,2012A&A...540A..85P}.  From 2400 K to 3100 K and log(g) from 4.0 to 5.5, [3.3$\micron$]-WISE1=0.01 with a range of $<$0.01.  Therefore, 2M1207 A is 11.60$\pm$0.03 and 2M1207 b is 15.46$\pm$0.10 in the [3.3$\micron$] filter.  Using a weighted average of parallax measurements \citep[52.8pc$\pm$1.0pc; ]{2007ApJ...669L..41B,2007ApJ...669L..45G,2008AA...477L...1D}, we find that 2M1207 b has an absolute [3.3$\micron$] magnitude of 11.85$\pm$0.14.  A summary of our photometry is presented in Table \ref{LBT/Magellan}.

\section{Mid-Infrared Colors of Exoplanets\label{mircmd}}
While there are still relatively few examples of directly imaged exoplanets, hundreds of brown dwarfs have been observed throughout the optical and infrared, and their properties are relatively well characterized compared to exoplanets.  Because brown dwarfs and gas-giants share similar compositions, temperatures, and radii, the large sample of brown dwarfs is a baseline with which to compare directly imaged exoplanets.

\begin{figure}
\begin{center}
\includegraphics[angle=90,width=\columnwidth]{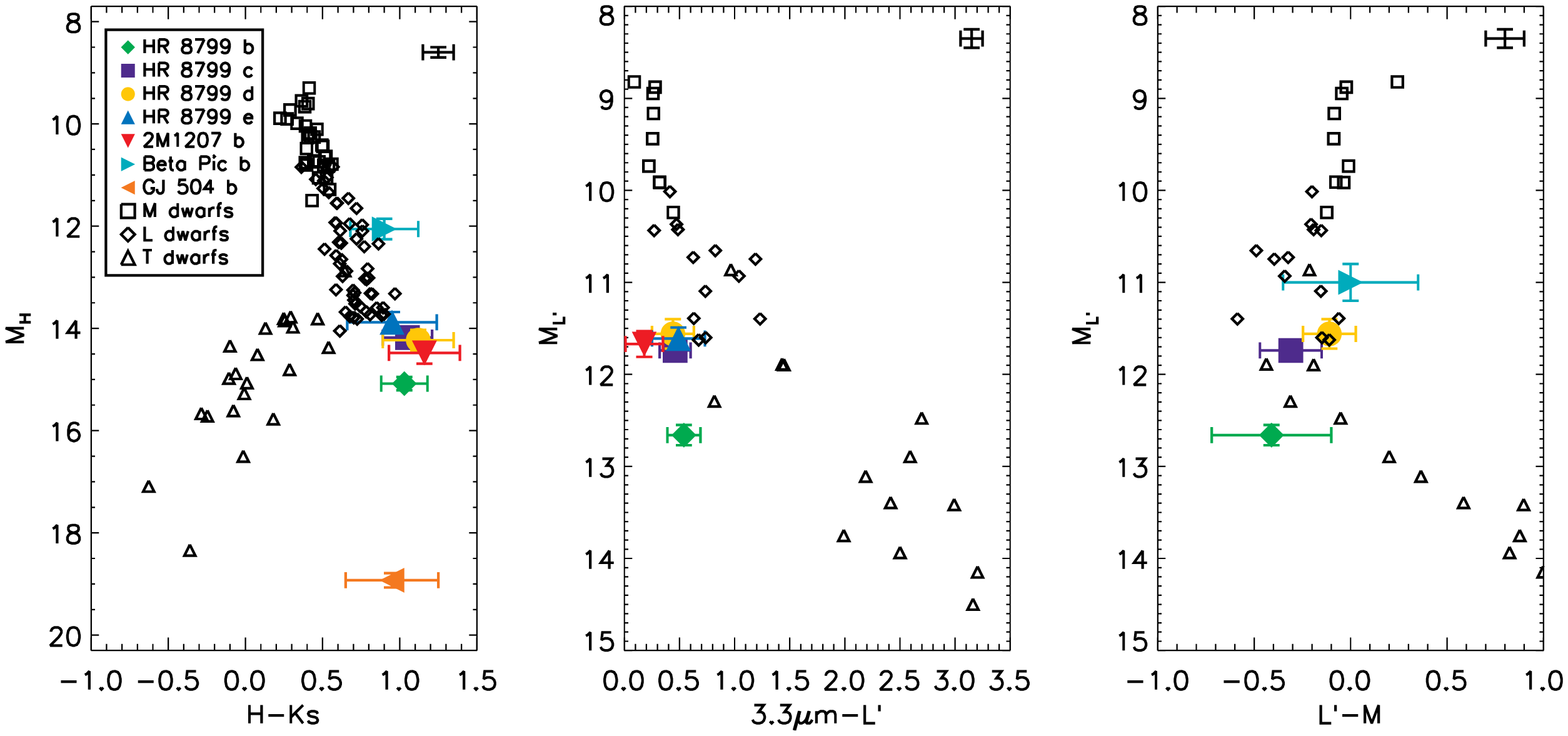}
\caption{Color-magnitude diagrams showing the positions of directly imaged extrasolar planets with respect to field brown dwarfs.  In the left-hand panel, the HR 8799 planets and 2M1207 b appear to be a continuation of the L dwarf sequence, implying that the planets have clouds at fainter absolute H-band magnitudes than the field brown dwarfs.  The recently discovered GJ 504 b also appears below the L dwarf sequence, although other near-infrared colors place it with the field T dwarfs.  In the center panel, the directly imaged planets are systematically bluer than the field brown dwarfs, suggesting that their 3.3$\micron$ methane absorption feature is somewhat diminished through some combination of clouds and non-equilibrium CO$\leftrightarrow$CH$_{4}$ chemistry.  The right-hand panel shows the directly-imaged planets as relatively consistent with the field brown dwarfs, although with large error bars.  For all three plots, error bars in the upper-right corner represent the maximum error bar for the brown dwarf sample.
\label{cmd}}
\end{center}
\end{figure}

In Figure \ref{cmd} we plot color-magnitude diagrams of brown dwarfs and directly-imaged exoplanets to illustrate similarities and differences in their appearances.  The brown dwarf data is assembled from \citet{2012ApJS..201...19D}, who include 2MASS photometry \citep{2006AJ....131.1163S}, WISE photometry \citep{2010AJ....140.1868W} and, when available, L' photometry \citep{1998ApJ...509..836L,2001ApJ...548..908L,2002ApJ...564..452L,2004AJ....127.3516G,2007ApJ...655.1079L,2010MNRAS.408L..56L} for a large sample of M, L and T-type brown dwarfs with parallax measurements.  We use \citet{2012ApJS..201...19D}'s set of ``normal'' brown dwarfs (excluding low-gravity, moving group members, and low-metallicity), removing a small fraction that have large photometric errors ($>$0.1 in color).  For the center and right color-magnitude plots ($\rm M_{L'}$ vs. [3.3$\micron$]-L' and $\rm M_{M}$ vs. L'-M) we use WISE band 1 and 2 measurements for the brown dwarf sequence and color correct\footnote{The color correction comes from fitting the complete set of BT-SETTL models \citep{2012RSPTA.370.2765A}, which cover a wide range of temperatures, and hence fully sample 3.3-WISE1 and M-WISE2 slopes.  We find that $[3.3\micron]-L'=-0.001+0.885\times(WISE1-L')+0.462567\times(WISE1-L')^2$ over a range of [0.0,2.0] in WISE1-L' (with corresponding polynomial errors of 0.003, 0.016 and 0.009), and $L-M'=-0.020+1.177\times(L'-WISE2)-0.002\times(L'-WISE2)^2$ over a range of [-1.0,1.5] in L'-WISE2 (with corresponding polynomial errors of (0.002, 0.008, and 0.007).} them to [3.3$\micron$] and M respectively so that they can be compared to ground-based measurements of exoplanets.  For the exoplanets, we use photometry of the four HR 8799 planets \citep{2008Sci...322.1348M,2010Natur.468.1080M,2011ApJ...739L..41G,2012ApJ...753...14S}, 2M1207 b \citep[][this work]{2004A&A...425L..29C}, $\beta$ Pic b \citep{2011A&A...528L..15B,2013A&A...555A.107B}, and GJ 504 b \citep{2013ApJ...774...11K}.  The exoplanets show a distinct separation from the field brown dwarfs in the left and center color-magnitude diagrams ($\rm M_{H}$ vs. H-Ks and $\rm M_{L'}$ vs. [3.3$\micron$]-L').

Brown dwarfs cool through the M dwarf and L dwarf sequence, becoming fainter and redder in the $\rm M_{H}$ vs. H-Ks color-magnitude diagram (left panel of Figure \ref{cmd}).  As the brown dwarfs cool below an absolute H magnitude of $\sim$14, the sequence moves sharply to bluer colors where the brown dwarfs have a spectral type of `T'.  The L$\rightarrow$T transition is thought to reflect the change from cloudy to cloud-free atmospheres \citep[e.g.][]{2006ApJ...640.1063B,2008ApJ...689.1327S,2010ApJ...723L.117M}, as the clouds dissipate (rain) and/or sink below the photosphere \citep{2001ApJ...556..872A,2002ApJ...571L.151B,2003ApJ...585L.151T,2004AJ....127.3553K}.  The planets in this range of absolute H magnitude appear to be an extension of the L dwarf sequence, implying that planets maintain cloudy photospheres at cooler effective temperatures than field brown dwarfs.  The distinction is thought to be gravity dependent\footnote{When comparing planets and brown dwarfs of the same effective temperature, the planets are younger and lower mass than the brown dwarfs.} \citep{2006ApJ...640.1063B,2008ApJ...689.1327S,2006ApJ...651.1166M,2010ApJ...723..850B,2013arXiv1310.0457L}.  In low-gravity objects, hydrostatic equilibrium implies that a given species' condensation temperature (where cloud decks form) will occur at a lower pressure (higher in the atmosphere) than in an equivalent temperature high-gravity object \citep{2012ApJ...754..135M}.  Thus the cloudy to cloud-free transition for low-gravity exoplanets should occur at lower effective temperatures and fainter H-magnitudes than for more massive brown dwarfs. For the exoplanets plotted in Figure 4a, $\beta$ Pic b \citep[$\sim$1700 K;][]{2013A&A...555A.107B}, HR 8799 cde \citep[$\sim$1100 K;][]{2011ApJ...729..128C,2012ApJ...753...14S}, 2M1207 b \citep[$\sim$1000 K;][]{2011ApJ...732..107S,2011ApJ...735L..39B} and HR 8799 b \citep[$\sim$900 K;][]{2011ApJ...729..128C} are all consistent with cloudy photospheres.  The recently discovered GJ 504 b \citep[$\sim$510;][]{2013ApJ...774...11K} is approximately the same H-Ks as the other planets while being much fainter overall.  However its J-H colors \citep[see Figure 11 of][]{2013ApJ...774...11K} are consistent with cloud-free, late T-type brown dwarfs, and narrow-band photometry shows methane absorption \citep{2013arXiv1310.4183J}.

At around the same temperature range where field brown dwarfs are transitioning from cloudy to cloud free, carbon is converted from CO to CH$_{4}$, producing strong methane absorption bands at 1.7 $\micron$ and 3.3$\micron$ \citep{1997ApJ...491..856B,2002ApJ...564..466G}.  For temperatures cooler than $\sim$1200K, virtually all of the carbon monoxide is removed if the reaction is allowed to go to chemical equilibrium \citep{2002Icar..155..393L,2011ApJ...735L..39B}.  However, CO absorption is observed in field T dwarfs \citep{1997ApJ...489L..87N,2009ApJ...695..844G}, implying that the reaction does not reach equilibrium faster than CO from hot brown dwarf interiors is mixed into their cooler photospheres \citep{2000ApJ...541..374S}.

While nearly all field brown dwarfs cooler than $\sim$1200K show CH$_{4}$ absorption in the near-infrared, exoplanets with similar effective temperatures do not \citep[HR 8799 bcd and 2M1207 b;][]{2007ApJ...657.1064M,2008Sci...322.1348M,2010ApJ...723..850B,2011ApJ...733...65B,2013Sci...339.1398K,2013ApJ...768...24O}.  In the mid-infrared, the CH$_{4}$ fundamental absorption band is at 3.3$\micron$ and a strong CO band is at 4.7$\micron$, while intermediate wavelengths ($\sim$4$\micron$) have relatively weak molecular opacity \citep{1997ApJ...491..856B}.  The center panel of Figure \ref{cmd} ($\rm M_{L'}$ vs. [3.3$\micron$]-L'), which is sensitive to CH$_{4}$ absorption, shows the field brown dwarfs becoming redder at later spectral types, due to increased methane absorption.  The planets (HR 8799 bcde and 2M1207 b) are systematically bluer than the field brown dwarfs, which could be caused by a lack of methane, which absorbs at 3.3$\micron$ and/or clouds, which smooth out spectral features (see Section \ref{SED modeling} for detailed modeling).  The right panel of Figure \ref{cmd} ($\rm M_{M}$ vs. L'-M), which is sensitive to CO absorption, also shows the brown dwarf sequence becoming redder at later spectral types, although in this case, the effect is caused by decreasing CO absorption.  The planets (HR 8799 bcd) are roughly consistent with the brown dwarf sequence (within their large error bars), indicating that they may have a similar non-equilibrium excess of CO.

\section{Spectral Energy Distributions and Modeling\label{SED modeling}}
The color-magnitude diagrams described in Section \ref{mircmd} suggest that exoplanets maintain cloudy photospheres at lower temperatures than field brown dwarfs, and that non-equilibrium CO$\leftrightarrow$CH$_{4}$ chemistry is at least partially responsible for their 3-5$\micron$ SEDs.  While there has been relative success fitting the near-infrared (1-2.5$\micron$) SEDs of these objects with cloudy and/or non-equilibrium chemistry models \citep{2008Sci...322.1348M,2011ApJ...729..128C,2011ApJ...737...34M,2011ApJ...733...65B,2011ApJ...732..107S,2011ApJ...735L..39B,2013Sci...339.1398K}, there has been more difficulty when the mid-infrared (3-5$\micron$) SEDs are considered in parallel \citep{2010ApJ...716..417H,2012ApJ...753...14S,2012ApJ...754..135M,2013arXiv1307.1404L}.  Here we attempt to explain the broad-wavelength SEDs of planets, incorporating our new photometry of HR 8799 c, HR 8799 d, and 2M1207 b. Sections 4.1-4.3 review models from previous work, in the context of our new data.  Section 4.4-4.5 present new models.  Section 4.6 summarizes our aggregate findings.  The models broadly fall under three families, which, for simplicity, we denote by their group leaders: \textit{Burrows}, \textit{Barman}, and \textit{Marley}.  A listing of all of the models and their basic properties is given in Table \ref{model listing}.  The photometry plotted in figures is listed in Tables \ref{LBT/Magellan} (new data) and \ref{lit phot} (literature data).

\subsection{Previous Work: Cloudy/Thick-Cloudy Atmospheres\label{thick-cloudy subsec}}
At the time of their initial discovery it was noticed that the HR 8799 planets had colors that looked like L dwarfs (cloudy with no methane absorption), while their luminosities were more consistent with field T dwarfs \citep{2008Sci...322.1348M}.  A similar phenomenon had been seen in the planetary-mass object, 2M1207 b.  But as a single object, it could be explained as a ``normal'' brown dwarf seen through an edge-on disk \citep{2007ApJ...657.1064M}.  

\begin{figure}
\begin{center}
\includegraphics[angle=90,width=\columnwidth]{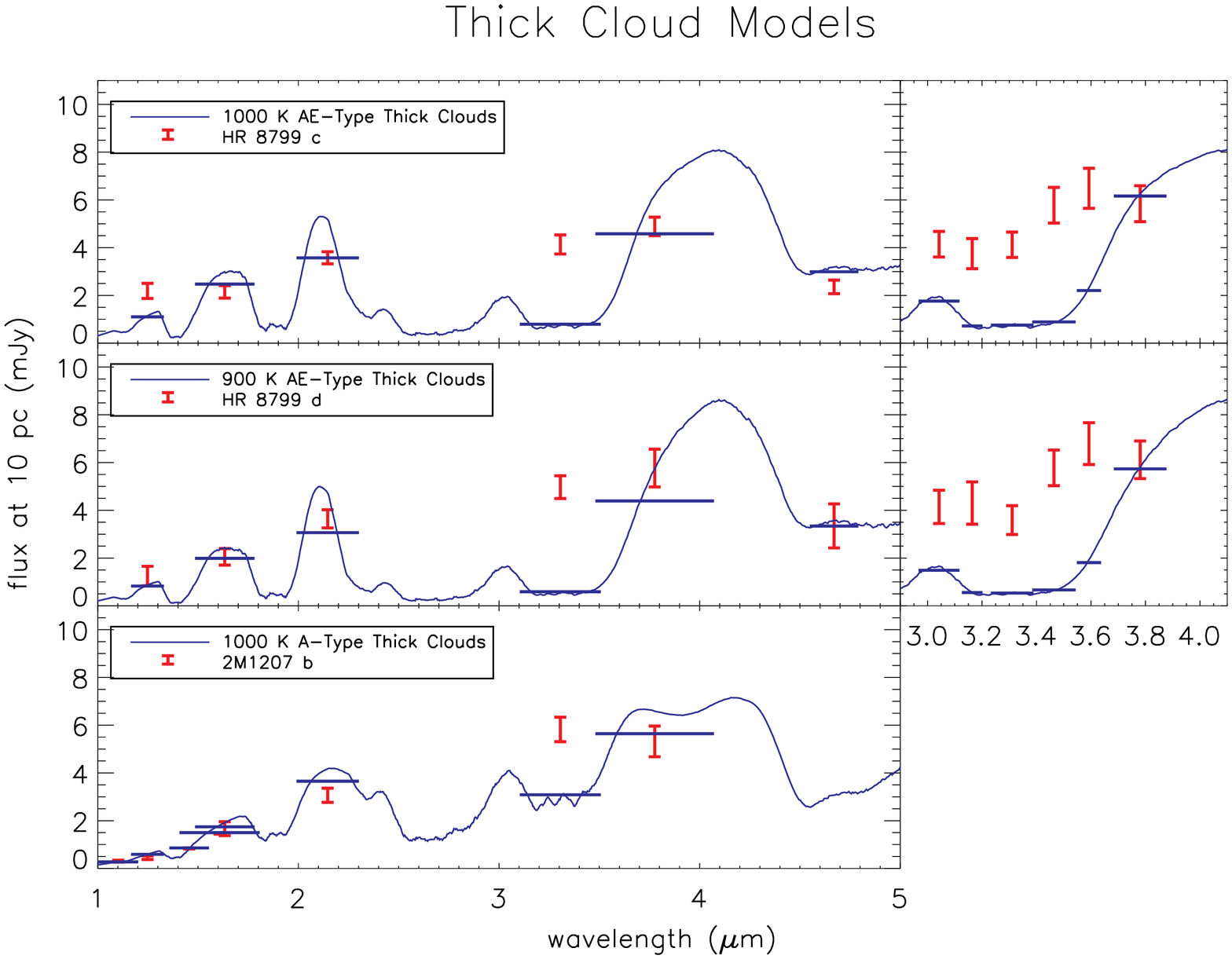}
\vspace{0.1in}
\caption[]{Absolute photometry for HR 8799 cd and 2M1207 b compared to thick-cloud atmosphere models.  The photometric data are shown as red error bars, the model spectra are shown as blue curves, and the model photometric predictions are shown as blue horizontal lines spanning the filter half-maximums.  The thick-cloud models are adopted from \citet{2011ApJ...737...34M} for HR 8799 cd and \citet{2011ApJ...732..107S} for 2M1207 b, using the grid from \citet{2011ApJ...737...34M}.  Two cloud parameterizations are used: A-type, which are the thickest clouds in \citet{2011ApJ...737...34M} and AE-type, which are slightly thinner.\\\\
For all three objects, the broad-band photometry (left-side of the figure) is generally well-fit by the models, except for the [3.3$\micron]$ filter.  Our new narrow-band 3-4$\micron$ photometry (right-side of the figure) is consistent with broader-band photometry at similar wavelengths, although systematically brighter within errors.
\label{thick clouds fig}}
\end{center}
\end{figure}

Since the HR 8799 planets are also redder than the field L dwarfs (see Figure \ref{cmd}), \citet{2011ApJ...729..128C} and \citet{2011ApJ...737...34M} used thick clouds to explain their broad SEDs, and \citet{2011ApJ...732..107S} used the same models to explain 2M1207 b.  The adopted models from \citet{2011ApJ...737...34M} and \citet{2011ApJ...732..107S} are shown in Figure \ref{thick clouds fig}.  For HR 8799 c and d, the models under-predict the [3.3$\micron$] flux by a factor of $\sim$4 \citep{2012ApJ...753...14S}.  

Our new narrowband photometry of HR 8799 c and d (shown in the right panel) are consistent with the broad-band photometry (in the left panel) and show a shallow slope with a slight break between L$_{NB3}$ and L$_{NB4}$ (which is consistent with methane's opacity profile; see Figure \ref{filter profiles}), rather than the deep methane absorption feature seen in the thick-cloud models.  Our new photometry of 2M1207 b shows that it is also much brighter at [3.3$\micron$] than the thick-cloud models predict (by a factor of$\sim$2, but this is very dependent on the precise choice of cloud thickness), and it has a similar, if not more extreme 3.3$\micron$-3.8$\micron$ slope as the HR 8799 planets.

\subsection{Previous Work: Cloudy Atmospheres with Non-Equilibrium CO$\leftrightarrow$CH$_{4}$ Chemistry}
After obtaining a near-infrared spectrum of HR 8799 b, \citet{2011ApJ...733...65B} showed that in addition to clouds, non-equilibrium CO$\leftrightarrow$CH$_{4}$ chemistry, caused by vertical mixing, is required to explain the spectrum of HR 8799 b.  Subsequently, \citet{2011ApJ...735L..39B} and \citet{2011ApJ...739L..41G} applied similar models to 2M1207 b, and HR 8799 c and d respectively.

\begin{figure}
\begin{center}
\includegraphics[angle=90,width=\columnwidth]{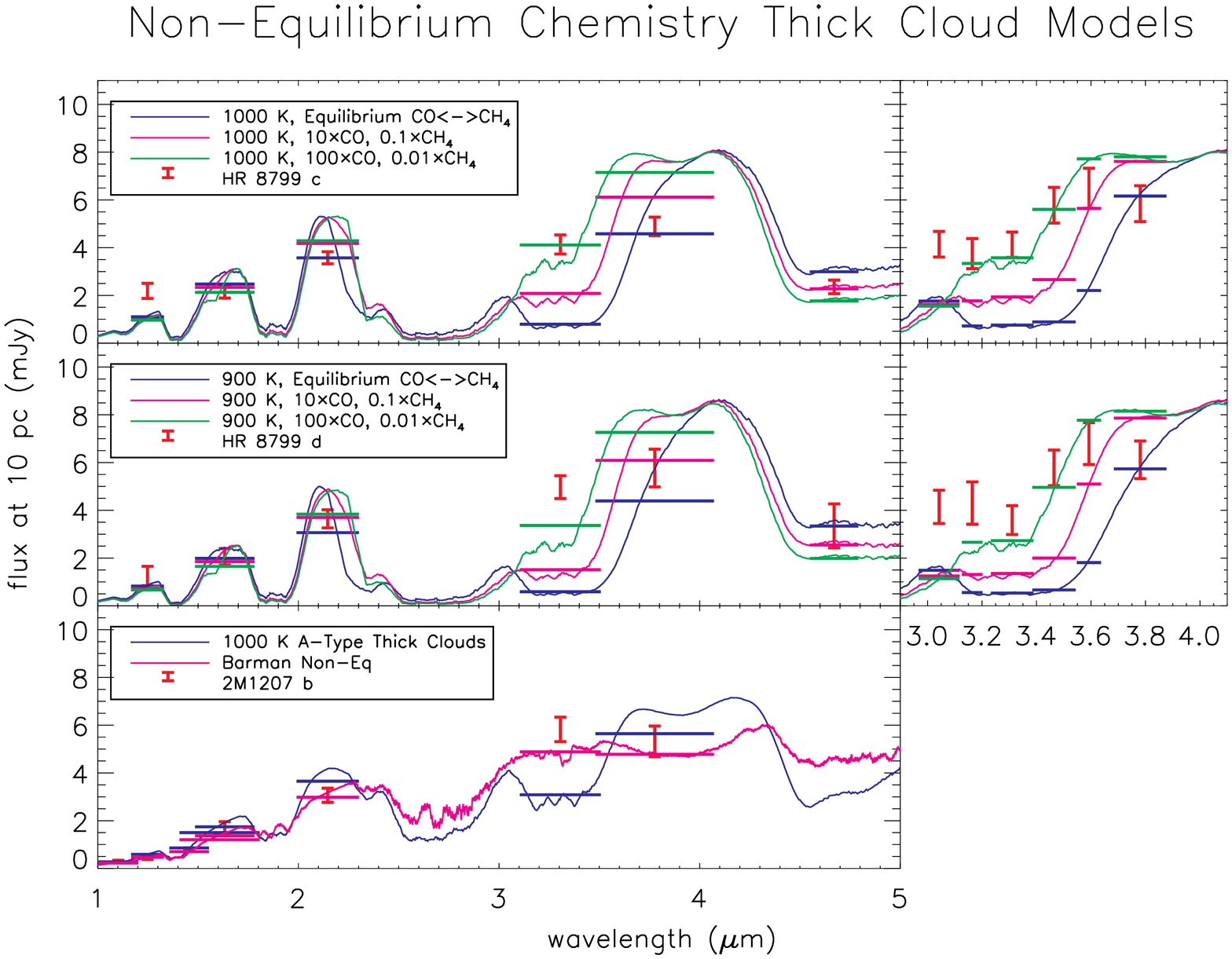}
\vspace{0.1in}
\caption[]{Same as Figure \ref{thick clouds fig} but with models that include non-equilibrium CO$\leftrightarrow$CH$_{4}$ chemistry or parameterized CO/CH$_{4}$ opacity.  The \textit{Burrows} models are from \citet{2012ApJ...753...14S} and have similar clouds properties as Figure \ref{thick clouds fig} but with diminished CH$_{4}$ and enhanced CO opacity.  The \textit{Barman} model, which is from \citet{2011ApJ...735L..39B} also employs thick clouds, and uses chemical reaction rates to self-consistently calculate CH$_{4}$ and CO mixing ratios assuming a vertical diffusion coefficient of $K_{zz}=10^{8}cm^{2}/s$.\\\\
The HR 8799 planets are not well-fit by the \textit{Burrows} models, which predict a sharp edge to the methane absorption feature, even with suppressed methane opacity.  Our new narrow-band 3-4$\micron$ photometry suggest a smoother slope.  Note that analogous models from  \citet{2011ApJ...733...65B} and \citet{2011ApJ...739L..41G} predict qualitatively similar behavior (when considering models that have radii consistent with evolutionary tracks).  However, the \textit{Barman} model fits 2M1207 b quite well, including our new [3.3$\micron$] photometry.  Combined with \citet{2011ApJ...735L..39B}'s fit of the 2M1207 b near-infrared spectrum, it appears that thick clouds and non-equilibrium chemistry can explain all of the existing data for 2M1207 b.
\label{non-eq thick clouds fig}}
\end{center}
\end{figure}

\citet{2012ApJ...753...14S} obtained [3.3$\micron$] photometry of all four HR 8799 planets, which were much brighter than the thick-cloudy/equilibrium chemistry atmosphere models (described in Section \ref{thick-cloudy subsec}), and somewhat brighter than the non-equilibrium models from \citet{2011ApJ...733...65B} and \citet{2011ApJ...739L..41G}.  To investigate the effects of non-equilibrium chemistry, \citet{2012ApJ...753...14S} used the thick cloud atmospheres from \citet{2011ApJ...737...34M} and removed CH$_{4}$ until the models were consistent with the [3.3$\micron$] photometry; however, this made the models too bright compared to existing L' (3.8$\micron$) photometry.  The non-equilibrium models from \citet{2012ApJ...753...14S} are shown in Figure \ref{non-eq thick clouds fig}.  These models show a sharp flux increase at the edge of the methane absorption feature, similar to what is seen in the equilibrium chemistry models, but shifted in wavelength.  This is inconsistent with the smooth slope seen in our narrowband photometry.

While the non-equilbrium chemistry models predict a fairly sharp increase at the edge of the methane absorption feature for HR 8799, the same is not true for 2M1207 b.  In Figure \ref{non-eq thick clouds fig}, we plot the \citet{2011ApJ...735L..39B} model of 2M1207 b along with our new [3.3$\micron$] photometry.  The photometry is consistent with the model.  2M1207 b is redder than the HR 8799 planets in the near-infrared.  This pushes the models towards thicker clouds, which wash out the methane absorption feature at 3.3$\micron$ in concert with non-equilibrium chemistry.  For HR 8799, the near-infrared driven cloud properties are not quite thick enough to wash out the 3-4$\micron$ SED, even with non-equilibrium chemistry.

\subsection{Previous Work: ``Patchy'' Linear Combinations of Cloudy Models}
\begin{figure}
\begin{center}
\includegraphics[angle=90,width=\columnwidth]{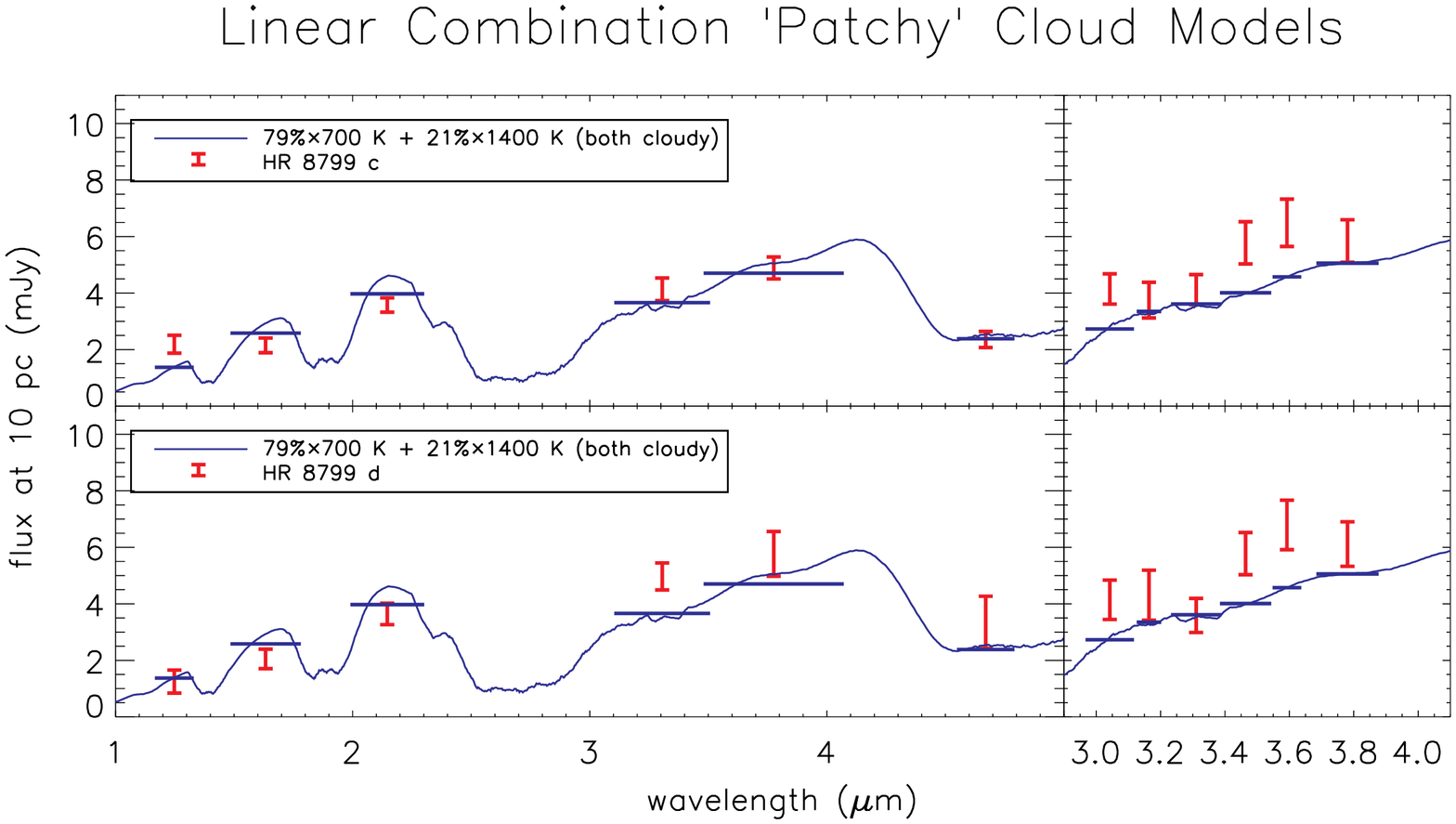}
\vspace{-1.3in}
\caption[]{Same as Figure \ref{thick clouds fig} but with models that linearly combine two different thick-cloud atmosphere models to represent non-isotropic emission through regions with different temperatures and different cloud properties.  The models are from \citet{2012ApJ...753...14S} using the grid from \citet{2011ApJ...737...34M}.\\\\
The HR 8799 planets are reasonably well-fit by the data.  However, the modeling approach of linearly combining two atmosphere models is not entirely self-consistent, so additional modeling (Section 4.4) is needed to study patchy cloud models.
\label{linear comb clouds fig}}
\end{center}
\end{figure}

Based on the earlier observations that 2M1207 b and the HR 8799 planets have L dwarf SEDs at T dwarf luminosities, \citet{2012ApJ...753...14S} attempted to fit the HR 8799 data with linear combinations of patchy cloud models, where the SED would be dominated by an L dwarf model, while a cooler, cloudier model suppressed its luminosity without drastically changing its shape.  The adopted models from \citet{2012ApJ...753...14S} are shown in Figure \ref{linear comb clouds fig}.  The combined models show an almost linear slope from 3-4$\micron$.  Our narrowband photometry has the same general shape as the models, but with, perhaps, a slight jump in flux between $L_{NB3}$ and $L_{NB4}$ (particularly for HR 8799 d), which would be better captured by the non-equilibrium chemistry models. Note that the narrowband 3-4$\micron$ photometry are systematically higher than the models (which were chosen to fit the broad-band photometry), but are consistent within error bars.

\subsection{Patchy Cloud Models (Cloudy/Cloudier and Cloudy/Cloud-Free)}
While the patchy cloud models from \citet{2012ApJ...753...14S} are encouraging, they are constructed by linearly combining a pair of one-dimensional atmospheric models which have different temperature profiles, including in the deep atmosphere where global circulation and convection would presumably homogenize their thermal profiles.  A self-consistent two column model needs to account for the energy flux carried by both components and should employ the same thermal profile.  Such a model is presented in \citet{2010ApJ...723L.117M} which allows for two atmosphere columns with arbitrary spatial coverage that combine to carry a specified thermal flux given by $\sigma T_{\rm eff}^4$. Using this formalism we construct illustrative models showing how heterogeneous clouds affect a planet's SED.  Figure \ref{patchy clouds fig} shows three models demonstrating the effect of self-consistently calculated cloud patchiness\footnote{The models are generally described in \citet{2008ApJ...689.1327S} and \citet{2012ApJ...754..135M}, with cloud model utilizing parameter $f_{\rm sed}$, from \citet{2001ApJ...556..872A} and \citet{2010ApJ...723L.117M}}: very thick (opaque) clouds ($f_{\rm sed}=0.25$), very thick clouds with patches of thinner clouds (same cloud model as the very thick clouds but with 10\% opacity), and very thick clouds with clear patches.  The models with no clear patches both wash out the near-infrared water absorption features \citep[the absorption is seen in the spectra of ][]{2011ApJ...733...65B,2013Sci...339.1398K} and suppress too much of the planet's near-infrared flux, pushing it to the mid-infrared. Small clear patches (here 10\% of the planet surface area) allow flux to escape from the holes and bring the near-infrared flux up to the measured values, but the resulting cooler atmosphere results in a large methane absorption feature at [3.3$\micron$], which is not seen in our mid-infrared photometry.  

\subsection{Patchy Cloud Models with Non-Equilibrium Chemistry\label{patchy non-eq subsec}}
\begin{figure}
\begin{center}
\includegraphics[angle=90,width=\columnwidth]{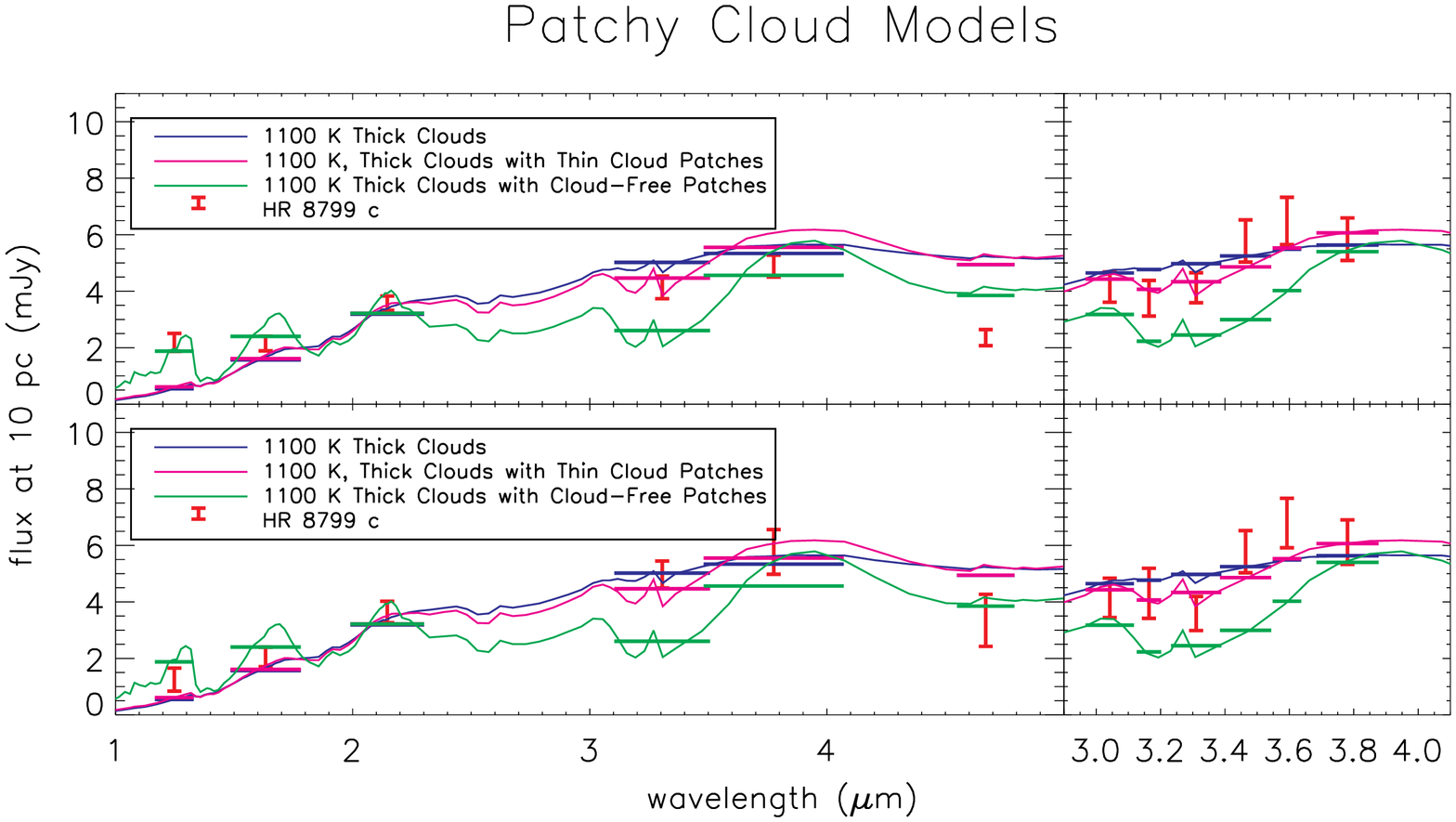}
\vspace{-1.3in}
\caption[]{Same as Figure \ref{thick clouds fig} but with self-consistent patchy clouds calculated using the formalism of \citet{2010ApJ...723L.117M}.  The thick cloud models use $f_{\rm sed}$=0.25 \citep{2001ApJ...556..872A}, which are very thick clouds.  The models with patches of thin clouds also use $f_{\rm sed}$=0.25, but with 30\% of the surface having 10\% of the normal cloud opacity.  In the third set of models, 10\% of the surface has no cloud opacity.\\\\
The HR 8799 planets are not well-fit by any of the models.  The thick cloud models and models with thin cloud patches both wash out the near-infrared SED where spectra of HR 8799 b, c, and 2M1207 b show strong water absorption bands \citep{2011ApJ...733...65B,2013Sci...339.1398K,2010A&A...517A..76P}.  The models with cloud-free holes still have too much methane absorption from $\sim$3.1-3.6$\micron$.
\label{patchy clouds fig}}
\end{center}
\end{figure}

\begin{figure}
\begin{center}
\includegraphics[angle=90,width=\columnwidth]{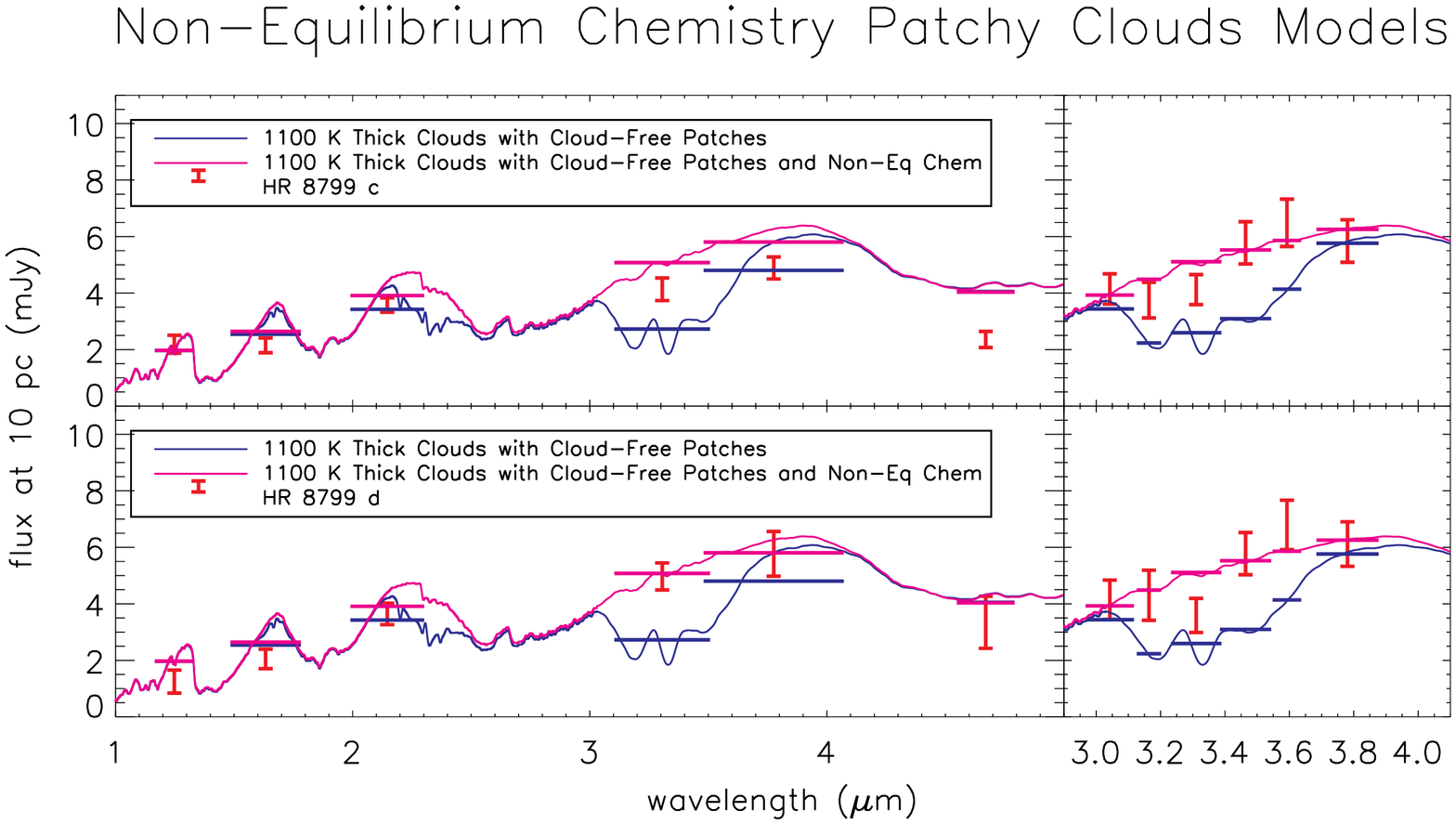}
\vspace{-1.3in}
\caption[]{Same as Figure \ref{thick clouds fig} but with self-consistent patchy clouds calculated using the formalism of \citet{2010ApJ...723L.117M} combined with non-equilibrium chemistry as described in \citet{2006ApJ...647..552S}.  The non-equilibrium models use a diffusion coefficient of $K_{zz}=10^{6}cm^{2}/s$.\\\\
Both of the HR 8799 planets are reasonably well-fit by the patchy-cloud/non-equilibrium chemistry model, although the model clearly over-predicts the M-band (4.7$\micron$) flux of HR 8799 c.  These models are a heuristic demonstration of the type of processes that are likely shaping the spectra of these directly imaged planets, although clearly more detailed model fits are required to find the best fitting parameters.
\label{non-eq patchy clouds fig}}
\end{center}
\end{figure}

In Figure \ref{patchy clouds fig} the $\sim$3.3$\micron$ absorption feature arises from relatively cool methane found very high in the atmosphere, at pressures of 10 to 20 mbar and temperatures of 875 to 900 K.  Because this is close to the $\rm CH_4 / CO$ equilibrium conditions and the deeper atmosphere is well within the region of CO stability, moderate vertical mixing can mix CO into the upper atmosphere and substantially lower the methane mixing ratio.  Figure \ref{non-eq patchy clouds fig} compares the cloudy/cloud-free model spectrum from Figure \ref{patchy clouds fig} with a spectrum computed following the prescription of \citet{2006ApJ...647..552S} for the same temperature and cloud profile but with eddy diffusivity $K_{zz} = 10^6\,\rm cm^2\, s^{-1}$.  With such mixing the methane number mixing ratio drops from $\sim 10^{-4}$ to $\sim 10^{-7}$ and the absorption feature is absent.  Such levels of mixing are consistent with those observed in some brown dwarfs \citep{2009ApJ...702..154S} and are broadly consistent with the strong gravity wave flux observed in numerical models of convection under these conditions \citep{2010A&A...513A..19F}.  The model spectra provide a qualitatively good fit to both HR 8799 c and d, although we have not attempted to fine-tune the models to fit the curvature seen in the narrowband filter SED, or constrain the exact model parameters.

\subsection{Summary of Modeling\label{model summary}}
For the HR 8799 planets, we find that a combination of patchy clouds and non-equilibrium CO$\leftrightarrow$CH$_{4}$ chemistry is necessary to fit the observed data.  Our 3-4$\micron$ photometry, in particular, is not fit well by models that employ patchy clouds without non-equilibrium chemistry, or non-equilibrium chemistry without patchy clouds.  For 2M1207 b, non-equilibrium chemistry models, which employ thick, but not patchy clouds, provide an adequate fit, and there is no reason to add extra free parameters.

This result suggests that all three planets have vertical mixing/non-equilibrium chemistry in their atmospheres, and all three planets have thick clouds, but only HR 8799 c and d necessarily have holes in their clouds.  It is possible that the HR 8799 planets are in the process of entering the L$\rightarrow$T transition, where cloud patchiness is observed as periodic variability in some field brown dwarfs \citep[e.g.,][]{2009ApJ...701.1534A,2012ApJ...750..105R}, while 2M1207 b, which is younger, has not yet evolved to that point.  As is the case for the brown dwarfs, the patchy clouds should cause variability in the HR 8799 planets.  Vertical mixing as evidenced by non-equilibrium chemistry appears to be persisting in the HR 8799 planets, even as the clouds are starting to dissipate.  For young, low-mass objects, the disappearance of clouds and the appearance of methane absorption might happen on different timescales and for different reasons.

Our work demonstrates that there are reasonable models that can explain the broad-wavelength appearances of the HR 8799 planets and 2M1207 b.  A dedicated modeling effort is needed to constrain the ranges of important physical parameters.

\section{Summary and Conclusions}
While the first generation of directly imaged planets were primarily discovered in the near-infrared (1-2.5$\micron$), cooler planets emit the majority of their light at longer wavelengths, where it is therefore critical to study their general appearances.  In particular, the 3-5$\micron$ region contains a significant amount of information about gaseous and condensed absorbers, while also being accessible to current ground-based telescopes.  

In this work, we have obtained photometry of HR 8799 c and d, and 2M1207 b in the 3-4$\micron$ window, which contains the methane fundamental absorption feature, and is a region where previous work has struggled to explain planet SEDs with atmospheric models.  The directly-imaged exoplanets are separated from the field brown dwarf population in $\rm M_{H}$ vs. H-K and $\rm M_{L''}$ vs. [3.3$\micron$]-L' color-magnitude diagrams, suggesting that the exoplanets are characteristically different than similar temperature, but older and more massive field brown dwarfs.  

Based on new and existing photometry, we attempt to model the complete SEDs of HR 8799 c and d, and 2M1207 b.  While models with thick clouds and vertical-mixing/non-equilibrium chemistry can explain the appearance of 2M1207 b, an additional free parameter for cloud patchiness is necessary to fit HR 8799 c and d.  Cloud patchiness is observed in the atmosphere of Jupiter and may signify the onset of the L$\rightarrow$T transition, where clouds dissipate, and/or sink below the photosphere of a planet/brown dwarf. In this scenario, the HR 8799 planets might be beginning their L$\rightarrow$T transition, while 2M1207 b is still too young.  Patchy clouds manifest as periodic variability in some brown dwarfs.  Our modeling predicts that HR 8799 c and d should be variable, although the magnitude of this result depends on the scale of the clouds holes and the orientation of the planets with respect to our line of sight \citep{2013ApJ...762...47K}.  Based on Figure 8, the variability should be evident at all wavelengths, with peak amplitudes in the J-band, and the short-wavelength side of the L-band.

The fact that the HR 8799 planets and 2M1207 b have different colors than field brown dwarfs suggests that it will be difficult to categorize brown dwarfs and extrasolar planets with sparse collections of infrared photometry.  Late-type brown dwarfs found by WISE, and ultra-cool exoplanets found by JWST imaging will both need to be studied in detail with multi-wavelength photometry and spectroscopy. Additionally, parallax measurements will be essential for breaking the degeneracy between nearby, young objects, and distant, older objects that might have similar colors.

Given that the planets studied in this work all have very shallow 3.3$\micron$-3.8$\micron$ slopes, it is worth considering whether L-band exoplanet imaging surveys should use a broader bandpass than the typically used L' filter (3.4-4.1$\micron$).  For HR 8799-like planets, a broader filter (3.0-4.1$\micron$) would improve sensitivity.  However, such a filter might not be a good match for cooler planets, if methane absorption becomes prominent at lower temperatures.

Currently, only a handful of exoplanets have been directly imaged.  As their numbers grow, it will be possible to determine how youth, gravity, composition and formation history affect their appearances.  And as the population of directly-imaged planets extends to cooler temperatures and lower masses, the mid-infrared will be a critical wavelength range for understanding their aggregate properties.

\acknowledgements
The authors thank Travis Barman for his insightful comments and for supplying his 2M1207 b model.  We also thank the anonymous referee for his/her excellent suggestions.  This work would not have been possible without the dedication of the LBTI staff, in particular Vidhya Vaitheeswaran, who programmed LBTI's rapid filter changing capabilities. AS was supported by the NASA Origins of Solar Systems Program, grant NNX13AJ17G. VB is supported by the NSF Graduate Research Fellowship Program (DGE-1143953).  The Large Binocular Telescope Interferometer is funded by the National Aeronautics and Space Administration as part of its Exoplanet Exploration program.  LMIRCam is funded by the National Science Foundation through grant NSF AST-0705296.

\clearpage

\begin{deluxetable}{lccccccccccc}
\tablecolumns{8}
\tabletypesize{\scriptsize}
\tablecaption{LBTI Photometry of HR 8799 c and d and Clio Photometry of 2M1207 b}
\tablewidth{0pt}
\tablehead{
\colhead{} &
\colhead{L$_{NB1}$} &
\colhead{L$_{NB2}$} &
\colhead{L$_{NB3}$} &
\colhead{L$_{NB4}$} &
\colhead{L$_{NB5}$} &
\colhead{L$_{NB6}$} &
\colhead{\[[3.3$\micron$]}
\\
\colhead{} &
\colhead{(3.04$\micron$)} &
\colhead{(3.16$\micron$)} &
\colhead{(3.31$\micron$)} &
\colhead{(3.46$\micron$)} &
\colhead{(3.59$\micron$)} &
\colhead{(3.78$\micron$)} &
\colhead{(3.31$\micron$)} 
}
\startdata
\sidehead{\textit{Contrast with Respect to Host Star}}
HR 8799 c & 10.12$\pm$0.15 & 10.17$\pm$0.20 & 9.97$\pm$0.15 & 9.52$\pm$0.15 & 9.32$\pm$0.15 & 9.32$\pm$0.15 & \\
HR 8799 d & 10.12$\pm$0.20 & 10.02$\pm$0.25 & 10.12$\pm$0.20 & 9.52$\pm$0.15 & 9.27$\pm$0.15 & 9.27$\pm$0.15 & \\
2M1207 b & & & & & & & 3.86$\pm$0.10 \\
\sidehead{\textit{Absolute Magnitudes}}
HR 8799 c & 12.35$\pm$0.15 & 12.40$\pm$0.20 & 12.20$\pm$0.15 & 11.75$\pm$0.15 & 11.55$\pm$0.15 & 11.55$\pm$0.15 & \\
HR 8799 d & 12.35$\pm$0.20 & 12.25$\pm$0.25 & 12.35$\pm$0.20 & 11.75$\pm$0.15 & 11.50$\pm$0.15 & 11.50$\pm$0.15 & \\
2M1207 b & & & & & & & 11.85$\pm$0.14 \\
\enddata
\tablecomments{The listed absolute magnitudes do not include a distance modulus uncertainty of 0.06 mag \citep{2007A&A...474..653V}, which is correlated between all bands.}
\label{LBT/Magellan}
\end{deluxetable}

\clearpage

\clearpage

\clearpage

\clearpage

\clearpage

\begin{deluxetable}{lcccccccccccc}
\tabletypesize{\scriptsize}
\tablecaption{Literature Photometry (Absolute Magnitudes)}
\tablewidth{0pt}
\tablehead{
\colhead{Filter} &
\colhead{$\lambda_{eff}$} &
\colhead{HR 8799 c} &
\colhead{HR 8799 d} &
\colhead{2M1207 b} &
\colhead{References}
}
\startdata
F090M            & 0.90$\micron$ &                &                & 18.83$\pm$0.25      & 1 \\
F110M            & 1.10$\micron$ &                &                & 16.98$\pm$0.15      & 1 \\
J                & 1.25$\micron$ & 14.65$\pm$0.17 & 15.26$\pm$0.43 & 16.37$\pm$0.20      & 2,2,3 \\
F145M            & 1.45$\micron$ &                &                & 15.42$\pm$0.03      & 1 \\
F160W            & 1.60$\micron$ &                &                & 14.63$\pm$0.02      & 1 \\
H                & 1.63$\micron$ & 14.18$\pm$0.14 & 14.23$\pm$0.2  & 14.46$\pm$0.21      & 4,4,5 \\
Ks               & 2.15$\micron$ & 13.13$\pm$0.08 & 13.11$\pm$0.12 & 13.30$\pm$0.11      & 2,2,5 \\
3.3              & 3.3$\micron$  & 12.2$\pm$0.11  & 12.0$\pm$0.11  &                     & 4,4 \\
L'               & 3.8$\micron$  & 11.74$\pm$0.09 & 11.56$\pm$0.16 & 11.65$\pm$0.14      & 2,2,5 \\
M                & 4.7$\micron$  & 12.05$\pm$0.14 & 11.67$\pm$0.35 &                     & 6,6 \\
\enddata

\tablerefs{
(1) \citet{2006ApJ...652..724S}
(2) \citet{2008Sci...322.1348M}
(3) \citet{2007ApJ...657.1064M}
(4) \citet{2012ApJ...753...14S}
(5) \citet{2004A&A...425L..29C}
(6) \citet{2011ApJ...739L..41G}
}

\label{lit phot}
\end{deluxetable}

\clearpage

\begin{deluxetable}{lcccccccccccc}
\rotate
\tabletypesize{\scriptsize}
\tablecaption{Atmospheric Models Used}
\tablewidth{0pt}
\tablehead{
\colhead{Figure} &
\colhead{Object} &
\colhead{Model Family} &
\colhead{T$_{eff}$} &
\colhead{Clouds} &
\colhead{Chemistry} &
\colhead{Reference}
}
\startdata
\ref{thick clouds fig} & HR 8799 c & Burrows & 1000 K & AE-type & Equilibrium & \citet{2011ApJ...737...34M} \\
\ref{thick clouds fig} & HR 8799 d & Burrows & 900 K  & AE-type & Equilibrium & \citet{2011ApJ...737...34M} \\
\ref{thick clouds fig} & 2M1207 b  & Burrows & 1000 K & A-type  & Equilibrium & \citet{2011ApJ...732..107S} \\
\hline
\ref{non-eq thick clouds fig} & HR 8799 c & Burrows & 1000 K & AE-type & Equilibrium & \citet{2011ApJ...737...34M} \\
\ref{non-eq thick clouds fig} & HR 8799 c & Burrows & 1000 K & AE-type & 0.1$\times$CH$_{4}$, 10$\times$CO & \citet{2012ApJ...753...14S} \\
\ref{non-eq thick clouds fig} & HR 8799 c & Burrows & 1000 K & AE-type & 0.01$\times$CH$_{4}$, 100$\times$CO & \citet{2012ApJ...753...14S} \\
\ref{non-eq thick clouds fig} & HR 8799 d & Burrows & 900 K & AE-type & Equilibrium & \citet{2011ApJ...737...34M} \\
\ref{non-eq thick clouds fig} & HR 8799 d & Burrows & 900 K & AE-type & 0.1$\times$CH$_{4}$, 10$\times$CO & \citet{2012ApJ...753...14S} \\
\ref{non-eq thick clouds fig} & HR 8799 d & Burrows & 900 K & AE-type & 0.01$\times$CH$_{4}$, 100$\times$CO & \citet{2012ApJ...753...14S} \\
\ref{non-eq thick clouds fig} & 2M1207 b  & Burrows & 1000 K & A-type  & Equilibrium & \citet{2011ApJ...732..107S} \\
\ref{non-eq thick clouds fig} & 2M1207 b  & Burrows & 1000 K & A-type  & Equilibrium & \citet{2011ApJ...732..107S} \\
\textbf{\ref{non-eq thick clouds fig}} & \textbf{2M1207 b}  & \textbf{Barman} & \textbf{1000 K} & \textbf{best-fit thickness}  & \textbf{Kzz}$\bm{=10^{8}}$ & \textbf{\citet{2011ApJ...735L..39B}} \\
\hline
\ref{linear comb clouds fig} & HR 8799 c & Burrows & 700K/1400K & A-type/AE-type & Equilibrium & \citet{2012ApJ...753...14S} \\
\ref{linear comb clouds fig} & HR 8799 d & Burrows & 700K/1400K & A-type/AE-type & Equilibrium & \citet{2012ApJ...753...14S} \\
\hline
\ref{patchy clouds fig} & HR 8799 c & Marley & 1100 K & $f_{\rm sed}$=0.25 & Equilibrium & this work \\
\ref{patchy clouds fig} & HR 8799 c & Marley & 1100 K & $f_{\rm sed}$=0.25 with 30\% thin clouds & Equilibrium & this work \\
\ref{patchy clouds fig} & HR 8799 c & Marley & 1100 K & $f_{\rm sed}$=0.25 with 10\% cloud free & Equilibrium & this work \\
\ref{patchy clouds fig} & HR 8799 d & Marley & 1100 K & $f_{\rm sed}$=0.25 & Equilibrium & this work \\
\ref{patchy clouds fig} & HR 8799 d & Marley & 1100 K & $f_{\rm sed}$=0.25 with 30\% thin clouds & Equilibrium & this work \\
\ref{patchy clouds fig} & HR 8799 d & Marley & 1100 K & $f_{\rm sed}$=0.25 with 10\% cloud free & Equilibrium & this work \\
\hline
\ref{non-eq patchy clouds fig} & HR 8799 c & Marley & 1100 K & $f_{\rm sed}$=0.25 with 10\% cloud free & Equilibrium & this work \\
\textbf{\ref{non-eq patchy clouds fig}} & \textbf{HR 8799 c} & \textbf{Marley} & \textbf{1100 K} & \textbf{$\bm{f_{\rm sed}}$=0.25 with 10\% cloud free} & \textbf{Kzz}$\bm{=10^{6}}$ & \textbf{this work} \\
\ref{non-eq patchy clouds fig} & HR 8799 d & Marley & 1100 K & $f_{\rm sed}$=0.25 with 10\% cloud free & Equilibrium & this work \\
\textbf{\ref{non-eq patchy clouds fig}} & \textbf{HR 8799 d} & \textbf{Marley} & \textbf{1100 K} & \textbf{$\bm{f_{\rm sed}}$=0.25 with 10\% cloud free} & \textbf{Kzz}$\bm{=10^{6}}$ & \textbf{this work} \\
\enddata
\tablecomments{The \textit{Burrows} models are generally described in \citet{1997ApJ...491..856B} and \citet{2006ApJ...640.1063B}, with cloud parameterizations from \citet{2011ApJ...737...34M}.  The \textit{Barman} models are comprehensively described in \citet{2011ApJ...733...65B} and references therein. The \textit{Marley} models are generally described in \citet{2008ApJ...689.1327S} and \citet{2012ApJ...754..135M}, with cloud parameterizations from \citet{2001ApJ...556..872A} and \citet{2010ApJ...723L.117M}.  Adopted models are highlighed in bold.}
\label{model listing}
\end{deluxetable}

\clearpage

\clearpage

\clearpage

\clearpage

\clearpage

\clearpage

\appendix
\section{Filter Properties}
\begin{figure}
\begin{center}
\includegraphics[angle=90,width=\columnwidth]{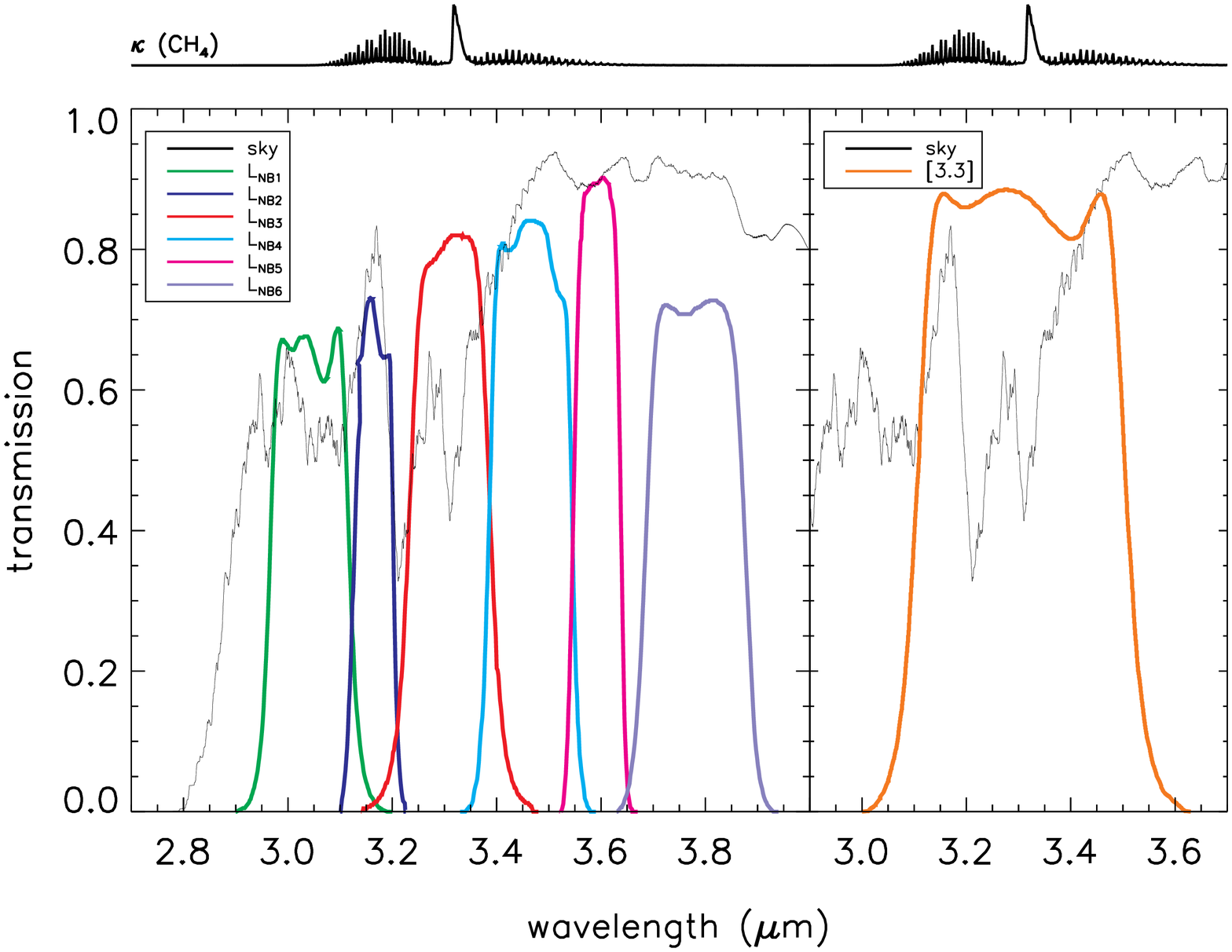}
\vspace{12pt}
\caption[]{Filter profiles for the seven filters used in this work and a model telluric profile for 1.0 airmasses and 4.3 mm precipitable water vapor\footnotemark.  The telluric transmission, which is variable, can significantly affect the raw filter properties, which are listed in Table \ref{filter table}.  Also shown is an opacity plot for methane (linear scaling), which demonstrates the utility of the filters for measuring methane absorption in cool atmospheres.
\label{filter profiles}}
\end{center}
\end{figure}

The observations described in this paper use non-standard filters in the 3-4$\micron$ L-band window.  We present the filter curves in Figure \ref{filter profiles} and calculate basic filter properties ($\lambda_{eff}$, FWHM, and zero-point flux) in Table \ref{filter table}.  The narrowband filters have not been used before, but are a useful set for measuring low-resolution SEDs in the L'-band window.  The [3.3$\micron$] filter has been used more extensively \citep{2010ApJ...716..417H,2011ApJ...729..128C,2012ApJ...753...14S} for sampling the short-wavelength side of the L-band, which is not covered by the more commonly used MKO L' filter.  Although this filter is colloquially referred to as L-short (Ls), it should not be confused with the L-short filter that was once installed in NIRC ($\lambda_{central}$=2.9785$\micron$, FWHM=1.023$\micron$)\footnote{http://www2.keck.hawaii.edu/inst/nirc/manual/fw\textunderscore macros.html}.  To avoid confusion, we refer to it as [3.3$\micron$] throughout this paper.

A model telluric atmosphere is shown with the filter set in Figure \ref{filter profiles}.  From 2.9-3.5$\micron$, the Earth's atmosphere has significant water-vapor absorption, which can vary significantly.  Sharp slopes in the telluric transmission can cause non-negligible changes to the filter properties listed in Table \ref{filter table} (in particular the 3.3$\micron$ filter), which were calculated without a telluric model.  Detailed modeling should include a telluric transmission model matching the observing conditions.

\footnotetext{Gemini Cerro Pachon model from http://www.gemini.edu/sciops/telescopes-and-sites/observing-condition-constraints/ir-transmission-spectra}
\clearpage

\begin{deluxetable}{lcccccccccccc}
\tabletypesize{\scriptsize}
\tablecaption{Filter Properties}
\tablewidth{0pt}
\tablehead{
\colhead{} &
\colhead{$\lambda_{eff}$ ($\micron$)} &
\colhead{FWHM ($\micron$)} &
\colhead{zero-point flux (Jy)}
}
\startdata
L$_{NB1}$                  & 3.04 & 0.15 & 361 \\
L$_{NB2}$                  & 3.16 & 0.08 & 342 \\
L$_{NB3}$                  & 3.31 & 0.16 & 313 \\
L$_{NB4}$                  & 3.46 & 0.16 & 290 \\
L$_{NB5}$                  & 3.59 & 0.09 & 270 \\
L$_{NB6}$                  & 3.78 & 0.19 & 243 \\
L$_{NB7}$\tablenotemark{a} & 3.95 & 0.24 & 225 \\
\[[3.3$\micron$]            & 3.31 & 0.40 & 311 \\
\enddata
\tablenotetext{a}{Not used in this work but available as part of the narrowband filter set.}
\tablecomments{$\lambda_{eff}$ and zero-point fluxes calculated using a synthetic Vega spectrum \citep{1992AJ....104.1650C}.  The filter properties are calculated without including telluric transmission, which varies with airmass and precipitable water vapor.}
\label{filter table}
\end{deluxetable}

\clearpage

\bibliographystyle{apj}
\bibliography{database}

\end{document}